\makeatletter
\declare@file@substitution{revtex4-1.cls}{revtex4-2.cls}
\makeatother

\documentclass{aastex631}

\usepackage{gensymb}
\usepackage{savesym}
\savesymbol{tablenum}
\usepackage{siunitx}
\restoresymbol{SIX}{tablenum}

\def\p{\partial}

\begin{document}

\title{Could a Bolide Listed in the CNEOS Database have Originated from 1I/'Oumuamua?}

\author[0000-0003-1116-576X]{Adam Hibberd}
\affiliation{Initiative for Interstellar Studies (i4is) 27/29 South Lambeth Road London, SW8 1SZ United Kingdom}








\begin{abstract}
The phenomenon of 1I/'Oumuamua introduced the interstellar object (ISO) class of celestial body into the astronomical lexicon, those objects with heliocentric speeds clearly in excess of that required to parabolically escape the Solar System - and therefore of extrasolar origin. A vogue topic at this moment in time is the possibility that some ISOs may impact with Earth, where they would be observed as bolides (meteor fireballs). There is the claim for instance that a meteor listed in the NASA-JPL CNEOS (Center for Near Earth Object Studies) database, CNEOS 2014-01-08 was interstellar, and additionally four further meteors from the database with interstellar origin have been proposed. This paper postulates that the origin of yet another meteor from this catalogue, CNEOS 2017-10-09 (observed over Bolivia, South America), was interstellar, as it may have been associated with 'Oumuamua. Note there is no direct velocity data for this object available, yet its observation time corresponds to the expected arrival time of an object ejected from 'Oumuamua and intersecting Earth's orbital position.  

\end{abstract}



\section{Introduction} \label{sec:intro}

With the arrival in the Solar System of interstellar object (ISO) now designated 1I/'Oumuamua, the first such object to be discovered, a new class of celestial body was formally established, with the designation letter \emph{'I'}, for interstellar.\\

The detection and relatively low perigee of 'Oumuamua ( $\sim{0.16}\ \si{au}$) naturally raised the possibility of ISOs colliding with the Earth, either in recent history \citep{Siraj_2022}, or stretching back over long periods of geological time \citep{llobet2022optimum}. The latter paper estimates an Earth-collision rate for 'Oumuamua-type objects of on average once every 10 million years. However it is reasonable to infer that smaller ISOs will strike more frequently.\\

This possibility led \cite{Siraj_2022} to study the NASA-JPL CNEOS database for bolides which might be interstellar meteors. Their search bore fruit in the form of object CNEOS 2014-01-08, which they identified as interstellar and further another object, CNEOS 2017-03-09. Both these objects are in the list of most likely bolides from CNEOS attributable as such in \cite{PAsensio_2022}, Table 8.\\

\begin{figure}[h]
\centering
\includegraphics[scale=0.57]{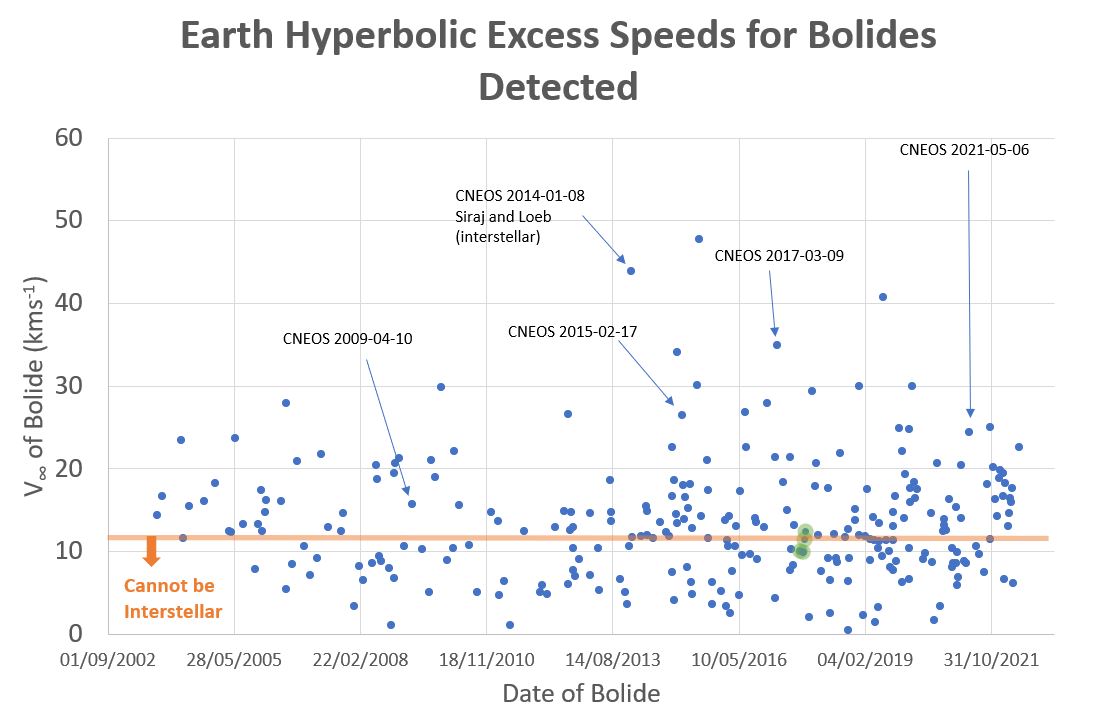}
\caption{Bolides up to early 2022 from NASA JPL CNEOS database for which we have sufficient data to compute their Earth hyperbolic excess speeds, error bars are unknown.}
\label{fig:bolides}
\end{figure}

For information, a chart of all bolides for which we have sufficient dynamical information to trace back their trajectories, is provided in Figure \ref{fig:bolides}. An in-depth analysis of these trajectories has already been performed by \cite{PAsensio_2022}.\\

A simple computation of the Earth hyperbolic excess speed of these bolides can discount certain of them as being interstellar (below the orange bar). Note that the uncertainty on this excess speed is unavailable, although \cite{BROWN201696,GRANVIK2018271} estimate an uncertainty in the velocities of CNEOS bolides as around 1 $\si{km.s^{-1}}$.\\

Figure \ref{fig:bolides} conveniently highlights all five bolides attributed by \cite{PAsensio_2022} as interstellar, with their respective CNEOS IDs, including the ones also identified by \cite{Siraj_2022}, CNEOS 2014-01-08 \& CNEOS 2017-03-09. The largest $V_{\infty}$ in this Figure belongs to CNEOS 2015-07-04 and is not mentioned in either \cite{Siraj_2022} or \cite{PAsensio_2022}, and it is assumed here that they have both discounted it as interstellar.\\

It should be emphasized that this claim by \cite{Siraj_2022} concerning CNEOS 2014-01-08 has stimulated a great deal of debate and there is also a fair degree of scepticism in the community of the likelihood of an interstellar origin for this object \citep{vaubaillon2022,Zuluaga_2019}, with \cite{HAJDUKOVA} articulating the complexity and uncertainty of such attributions to meteors in general.\\


In this paper, I address the possibility that a bolide detected in Bolivia, South America, on 9th October 2019 (henceforth CNEOS 2017-10-09) was interstellar. There is insufficient data on this body to determine its velocity of arrival and to thence trace back its heliocentric velocity in the manner exploited by Siraj and Loeb in their identification of meteor CNEOS 2014-01-08.\\ 

However, when one studies 'Oumuamua's trajectory, referencing back before its discovery date of 19th October 2017, and even before its perihelion on 9th September 2017, one finds an extremely low $\Delta V$ needed for an object leaving 'Oumuamua to eventually intercept Earth. Indeed this is evidenced in the subsequent orbital path followed by 'Oumuamua as it approached remarkably close to Earth, reaching a perigee as low as 0.16 $\si{au}$, an astronomical hair's breadth for an object originating from some remote location in our Milky Way galaxy.\\

\section{Analysis} \label{sec:anal}

First let us investigate the evolution of the minimum $\Delta V$ necessary to reach Earth from 'Oumuamua along its orbital path. \\

To establish this, \emph{Optimum Interplanetary Trajectory Software (OITS)} was utilised \citep{OITS_info}, in conjunction with the \emph{Non Linear Problem (NLP)} solving software NOMAD \citep{LeDigabel2011}.\\

\begin{figure}[h]
\centering
\includegraphics[scale=0.4150]{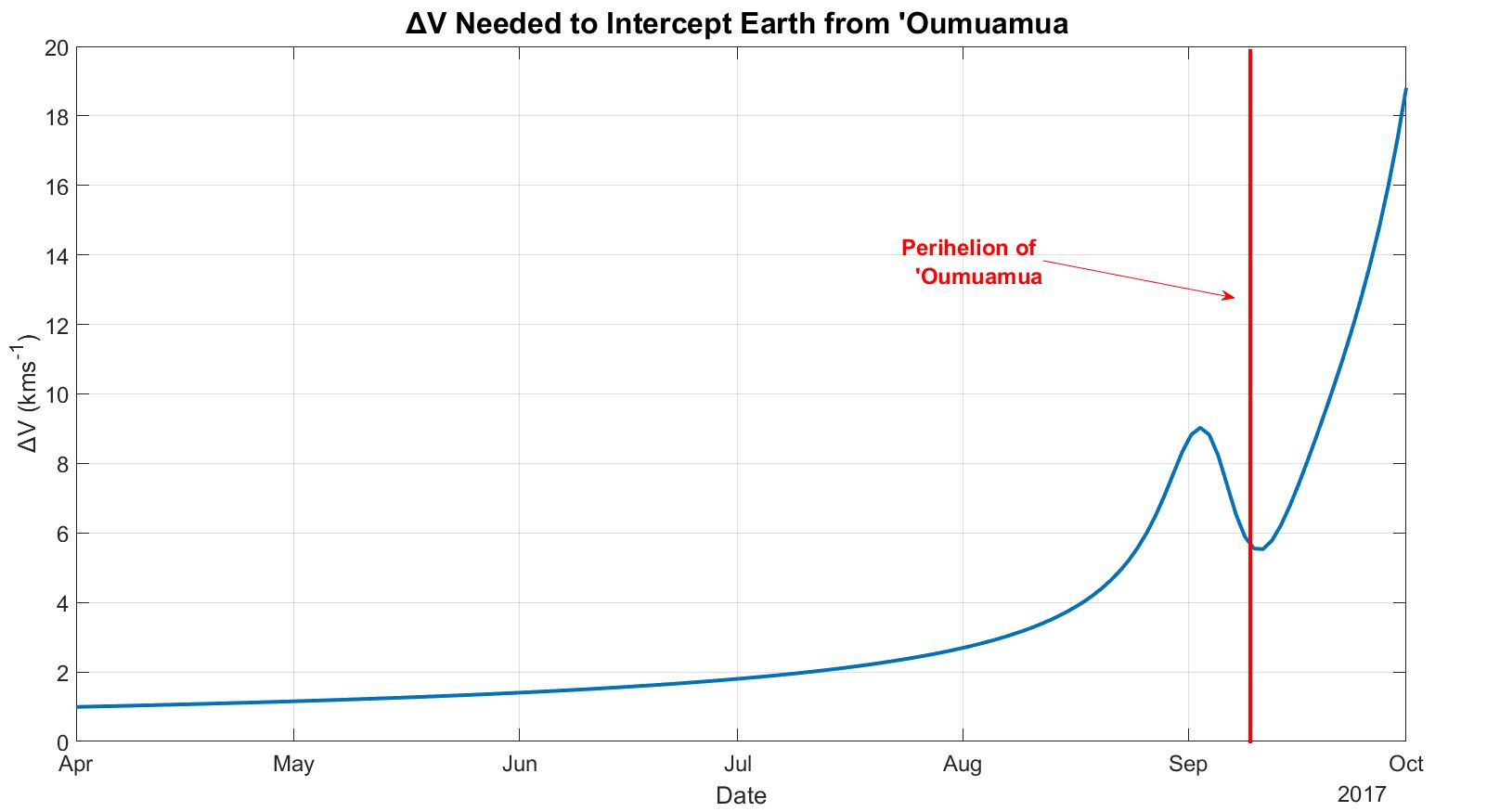}
\caption{Minimum $\Delta V$ required to reach Earth from interstellar object 'Oumuamua against time of departure}
\label{fig:mindv}
\end{figure}

Refer Figure \ref{fig:mindv} which reveals the results of this analysis. We find that the $\Delta V$ is always small before August 2017 ($<$ 3 $\si{km.s^{-1}}$) and is on a downward trend before April 2017 ($<$ 1 $\si{km.s^{-1}}$), whereas after August 2017 it begins to climb rapidly, despite a dip around perihelion, nevertheless continuing to escalate after this due to the increasingly imminent proximity of Earth, which 'Oumuamua flies by at a distance of only 0.16 \si{au}. \\

\begin{figure}[h]
\centering
\includegraphics[scale=0.4050]{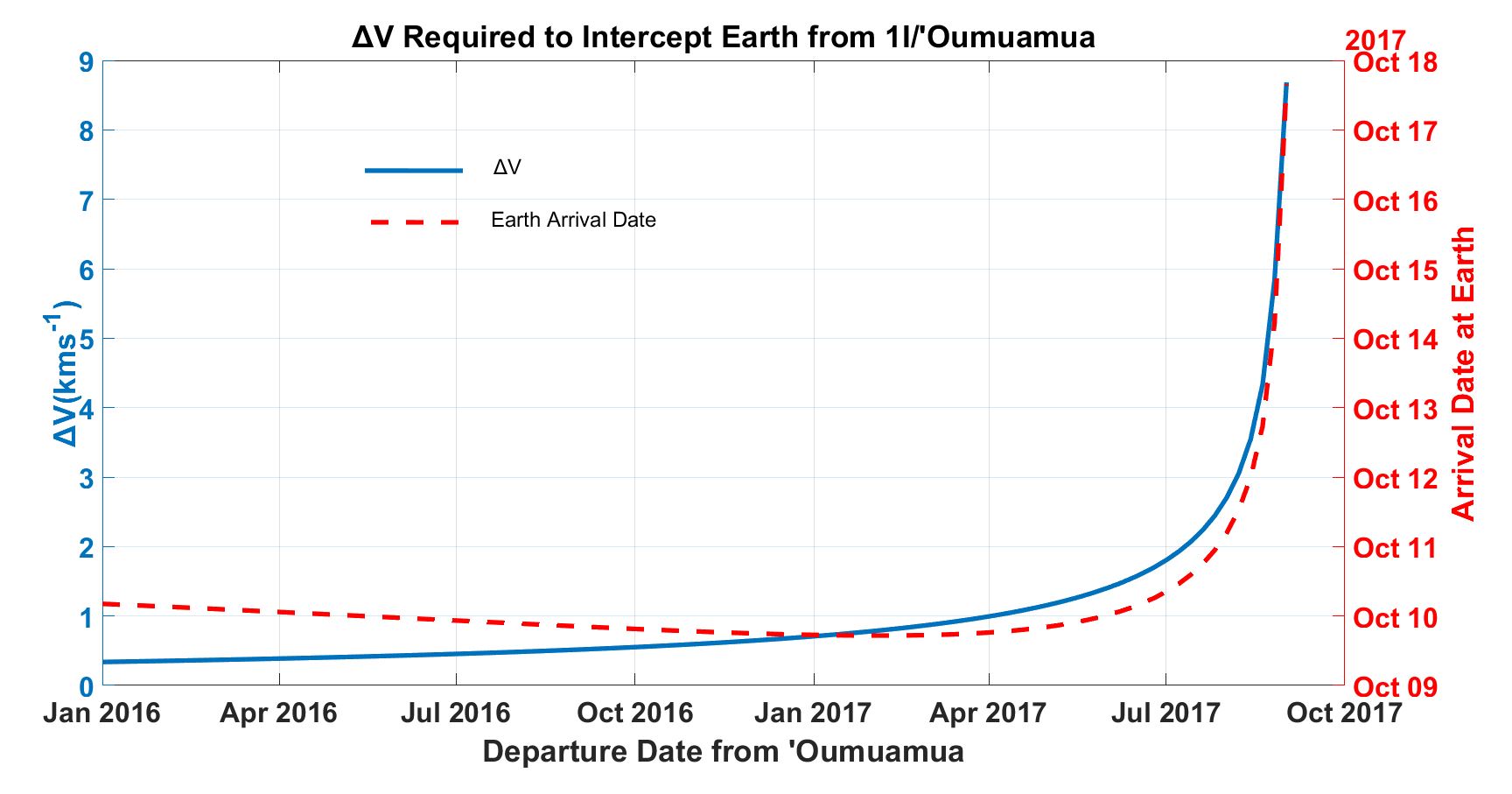}
\caption{Minimum $\Delta V$ required to reach Earth from interstellar object 'Oumuamua with time of Earth intercept (red-dashed curve and right-hand axis)}
\label{fig:mindvT}
\end{figure}

As well as extrapolating 'Oumuamua's orbital path backwards to an earlier time, Figure \ref{fig:mindvT} also provides (the red-dashed line) on the right-hand axis in red , the most likely arrival time at Earth for any object which has been ejected from 'Oumuamua. We discover that for a large portion of 'Oumuamua's trajectory, the most likely arrival date for any object ejected from 'Oumuamua would be 9th October 2017.\\

\begin{figure}[h]
\centering
\includegraphics[scale=0.400]{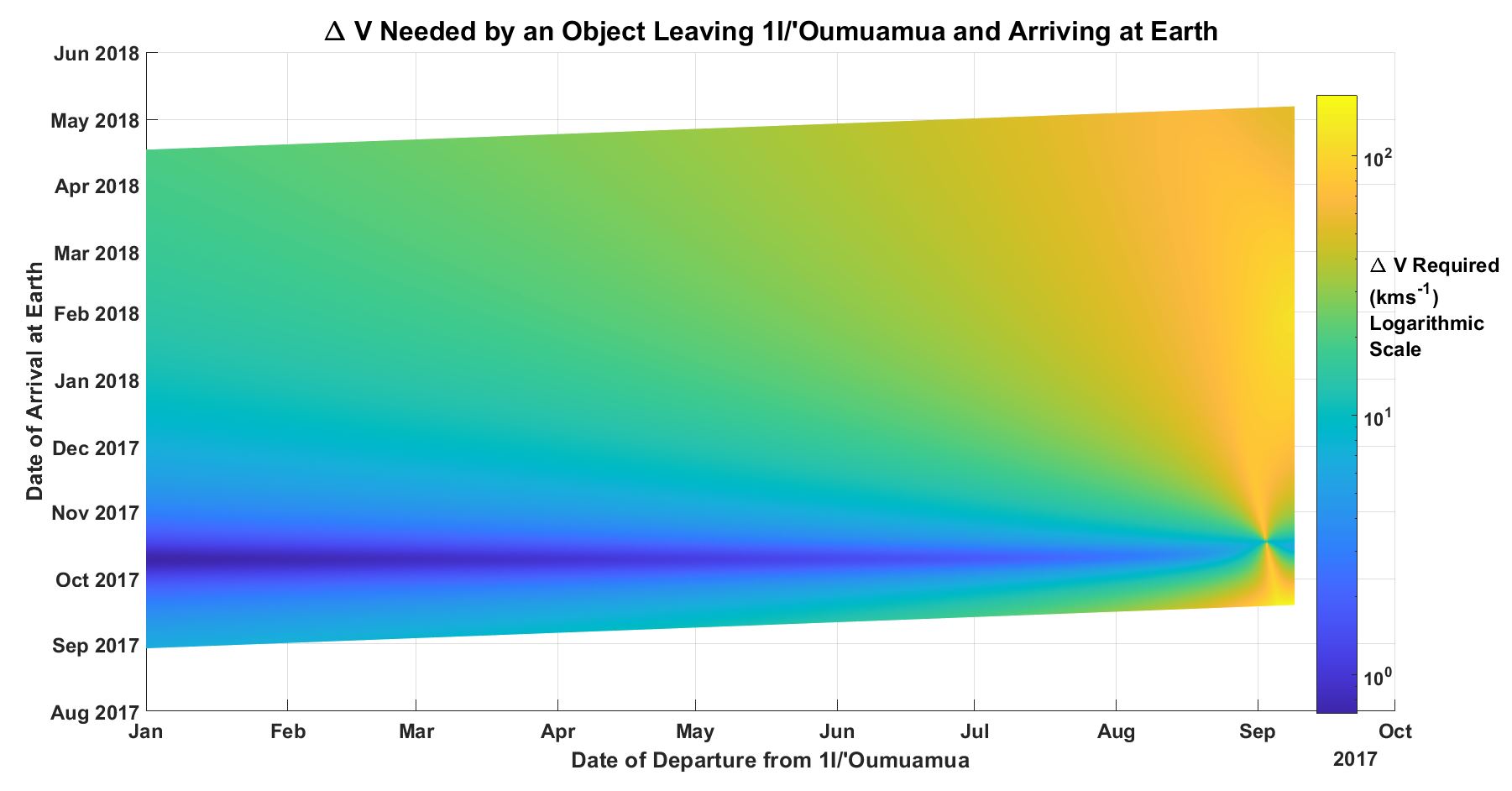}
\caption{Colour contour map of $\Delta V$ required to depart 'Oumuamua at time (x-axis) and arrive at Earth at time (y-axis), where the colour scale is logarithmic (see colour bar on right)}
\label{fig:ccmdv}
\end{figure}

This is confirmed in the colour contour map provided in Figure \ref{fig:ccmdv}, where the dark horizontal bar of particularly low $\Delta V$ values lies around the arrival dates 8th-12th October and becomes more dispersed with earlier departure date.\\

Reference to the CNEOS website allows access to data concerning bolides detected, starting way back around 1990. It is clearly pertinent to determine therefore whether there is an increase in the bolide detection rate around 8th-12th October 2017 corresponding to the most likely time-of-arrival for meteors with an origin of 'Oumuamua.\\ 

\begin{figure}[h]
\centering
\includegraphics[scale=0.4050]{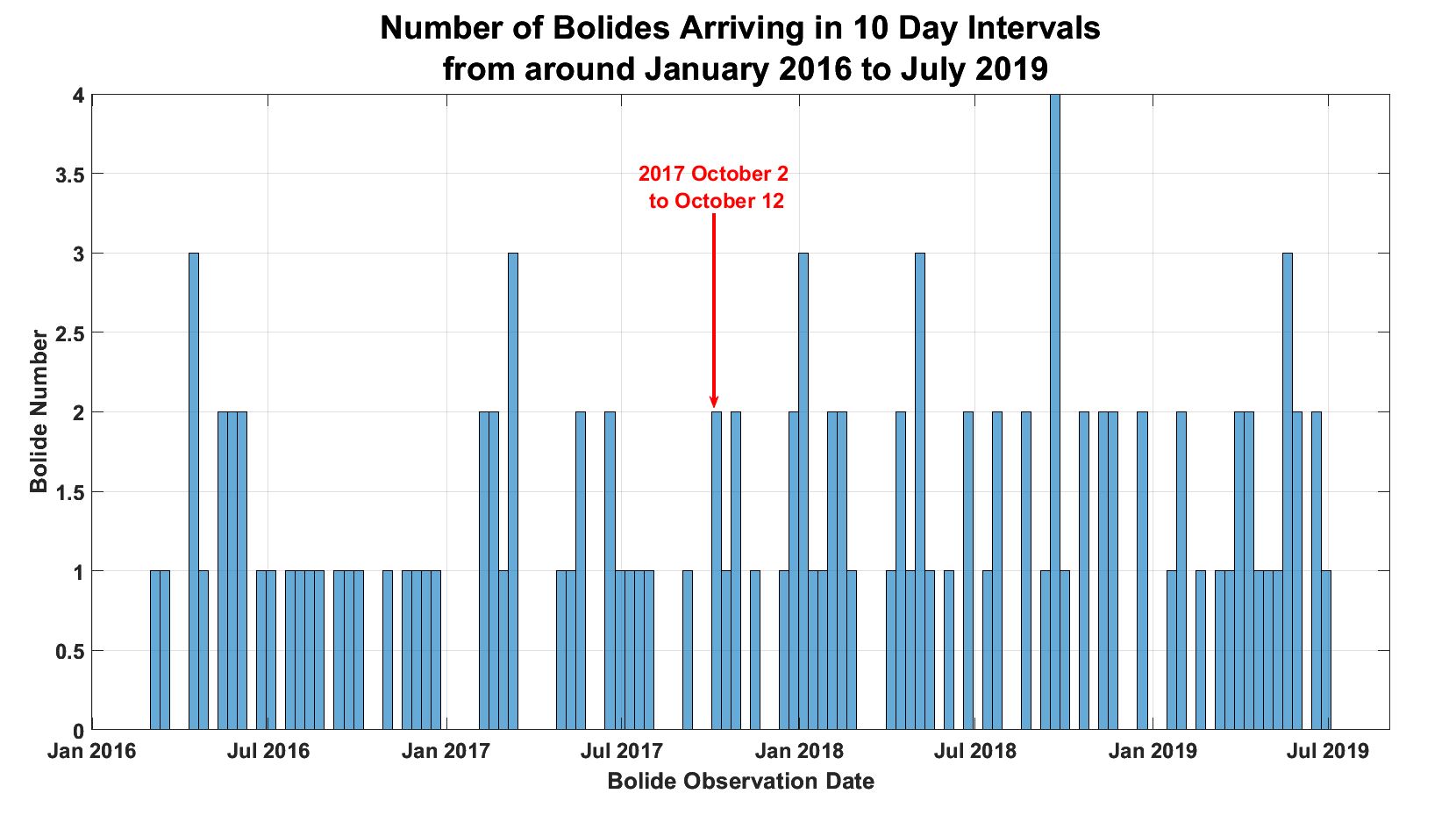}
\caption{Histogram of bolide arrival rate around the time of interest indicating NO particular increment on or around the expected arrival date (Derived CNEOS)}
\label{fig:bolidehist}
\end{figure}

To this end refer to Figure \ref{fig:bolidehist} which indicates no pronounced increase in bolide activity around this date. Nevertheless, there clearly ARE bolides which were observed around this time.\\

\begin{figure}[h]
\centering
\includegraphics[scale=0.4150]{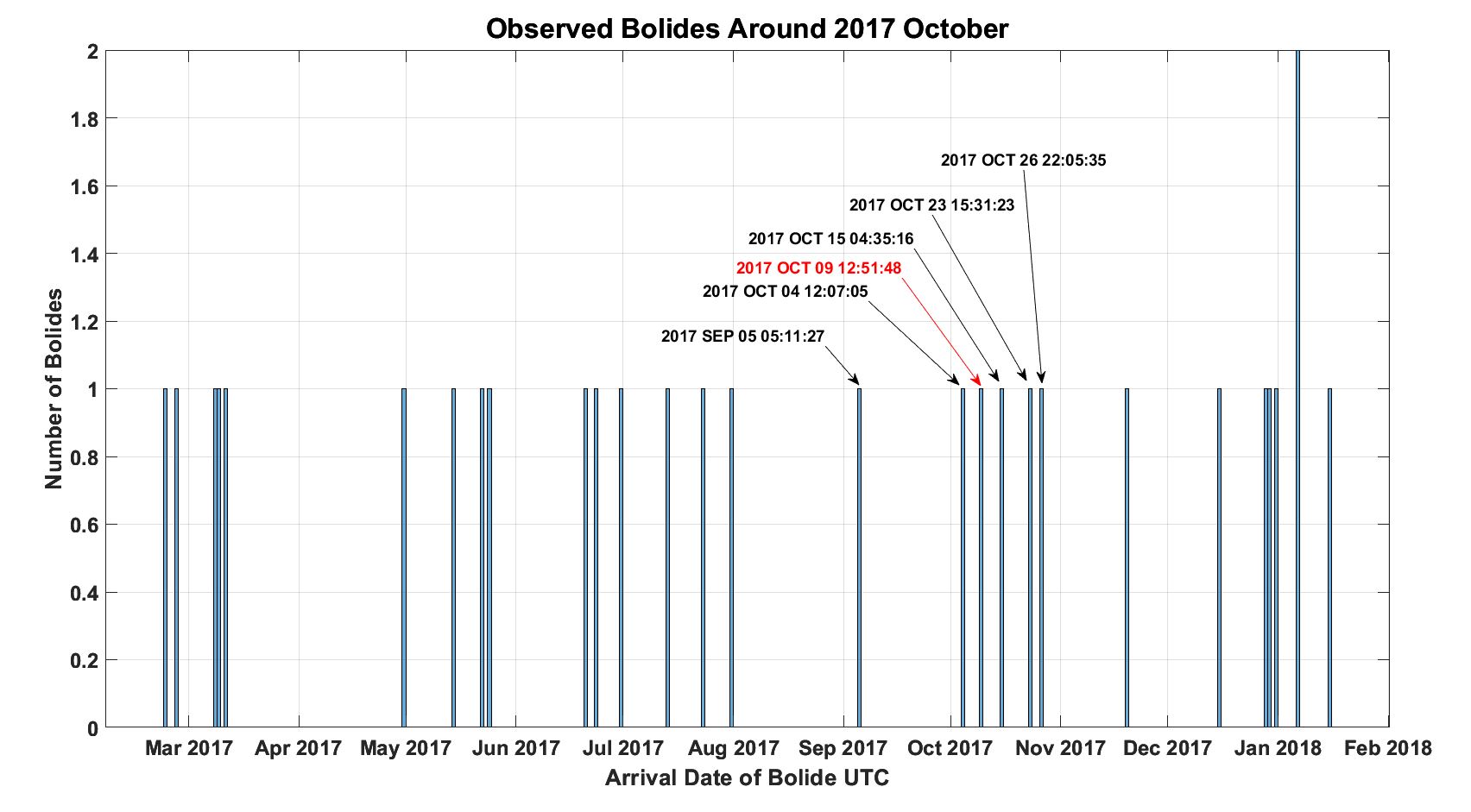}
\caption{Histogram of bolide arrival times (derived CNEOS) around the date of interest 2017 October 9th. The red highlighted bolide arrived on the precise date in question and is designated (CNEOS 2017-10-09).}
\label{fig:bolidezhist}
\end{figure}

\begin{figure}[h]
\centering
\includegraphics[scale=0.40]{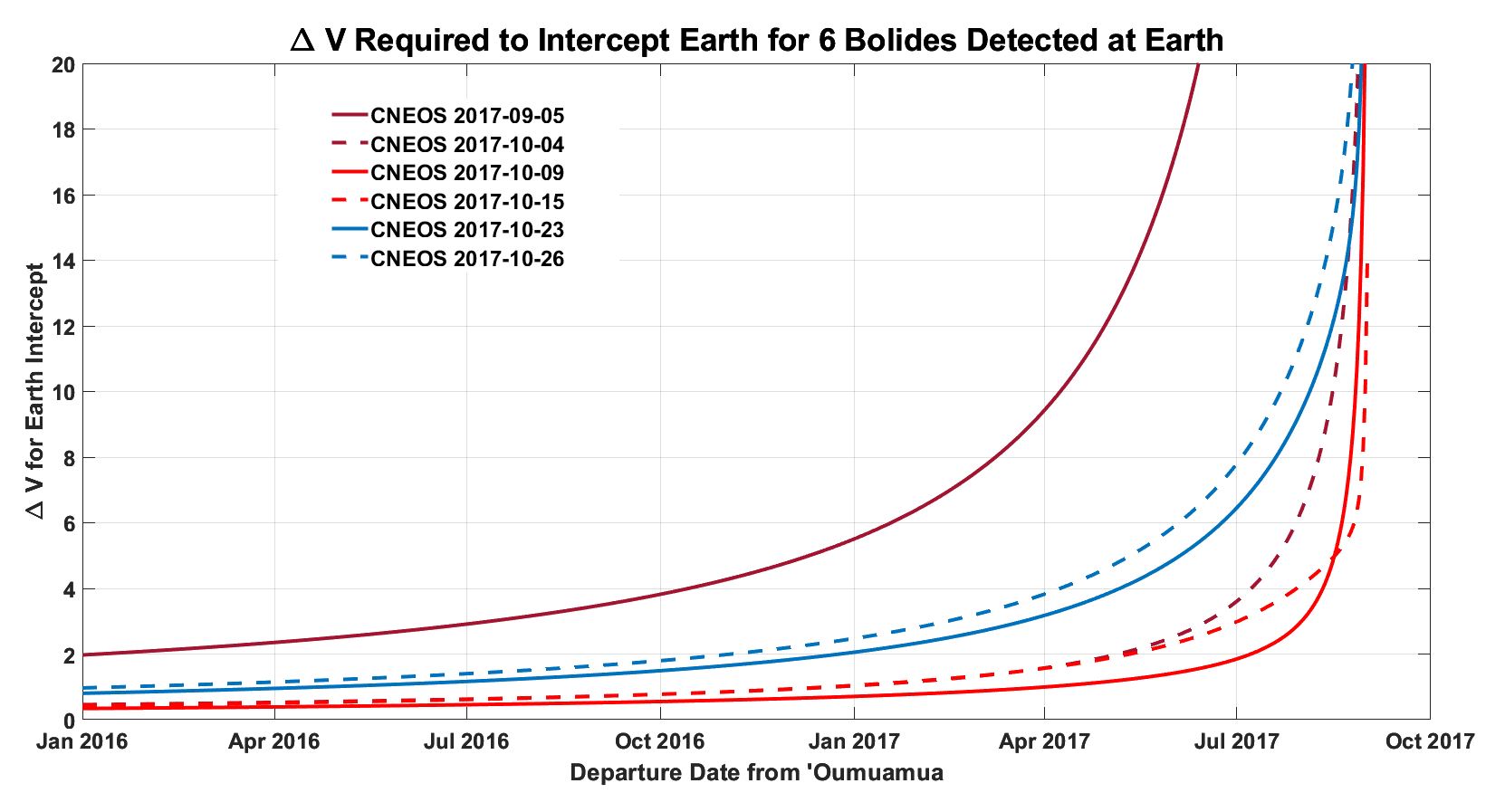}
\caption{$\Delta V$ needed for an object to be ejected from 'Oumuamua and arrive at Earth at the precise time of osbervation of 6 bolides from CNEOS 
which appeared on or around 2017-10-09, against possible time of ejection}
\label{fig:bolidedv}
\end{figure}

\begin{figure}[h]
\centering
\includegraphics[scale=0.47]{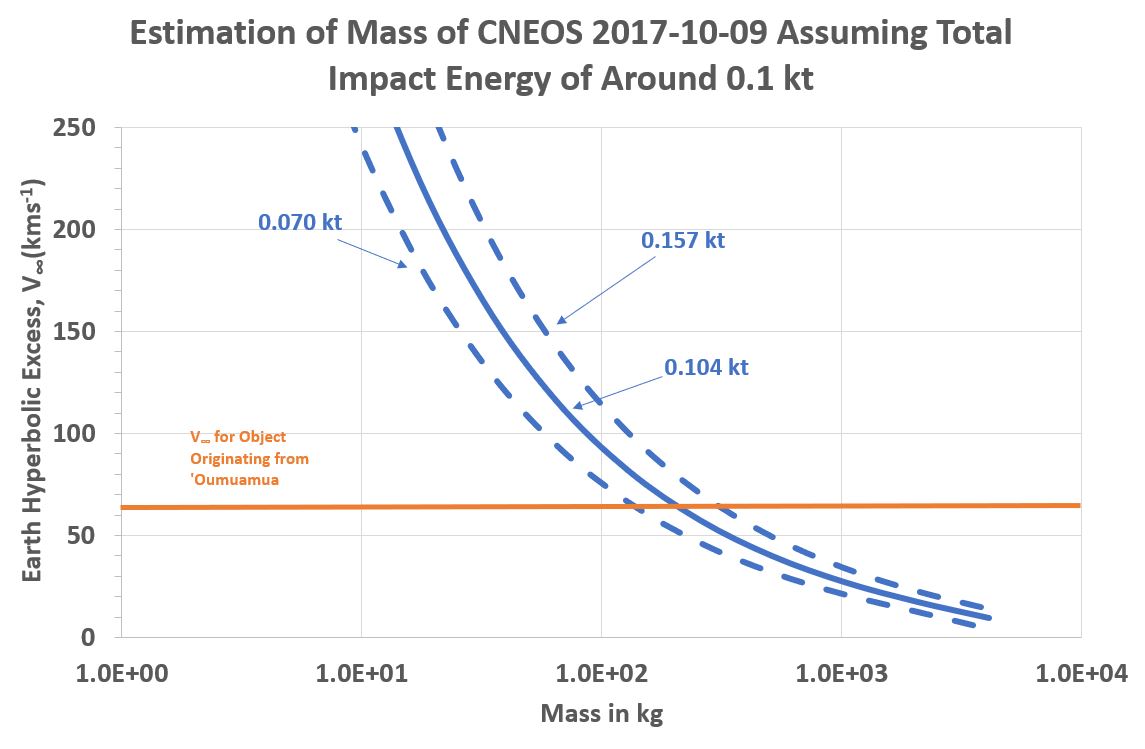}
\caption{Possible Earth hyperbolic excess speeds ($V_{\infty}$) of the CNEOS 2017-10-09 bolide against mass assuming a range of impact energies based on its measured radiative energy. The red line indicates the most likely $V_{\infty}$ for an object originating from 'Oumuamua.}
\label{fig:bolivian}
\end{figure}

Figure \ref{fig:bolidezhist} is a zoom in to around the dates in question and we find there are 6 bolides around 2017 October 9th, including the red highlighted bolide (CNEOS 2017-10-09) which arrived precisely on the expected arrival date. Of these 6 bolides, only 4 have sufficient state vector data from CNEOS to be included in Figure \ref{fig:bolides} - they are the green highlighted bolides towards the bottom right of Figure \ref{fig:bolides}, note that unfortunately, CNEOS 2017-10-09 is not amongst these 4, due to the absence of such data.\\

For information I provide in Figure \ref{fig:bolidedv}, for each of the 6 bolides discovered in the CNEOS around 2017-10-09, the $\Delta V$ needed to be ejected from 'Oumuamua against time of ejection such that Earth arrival occurs at the precise times UTC given in Figure \ref{fig:bolidezhist}.
\begin{table*}[]
\centering
\caption{CNEOS Bolides: Location and Local Time}
\label{tab:loctime}
\begin{tabular}{|c|c|c|c|c|c|c|c|}
\hline
\textbf{CNEOS} & \textbf{Time (UTC)} & \textbf{Time (Local)} & \textbf{Day} & \textbf{Location} & \textbf{$V_{\infty}$ (Earth)} & \textbf{$\Delta V$/Jan 2016} & \textbf{Latitude (\degree)}\\
 & & & \textbf{/Night} & &  \textbf{[\si{km.s^{-1}}]} & \textbf{[\si{km.s^{-1}}]} &\\ \hline
2017-09-05 & 05:11:27 & 22:22:27 & N & North Amer. & 10.0 & 1.98 & 49.3N\\
2017-10-04 & 12:07:05 & 20:07:05 & N & China & 9.9 & 0.46 & 28.1N\\
\textbf{2017-10-09} & 12:51:48 & 08:51:58 & \textbf{D} & Bolivia & N/A & 0.34 & 18.3S\\
\textbf{2017-10-15} & 04:33:16 & 13:33:16 & \textbf{D} & Antarctica & 11.3 & 0.46 & 65.2S\\
2017-10-23 & 15:31:23 & 18:31:23 & N & Saudi Arabia & 12.3 & 0.81 & 28.8N\\
\textbf{2017-10-26} & 22:05:35 & 10:05:35 & \textbf{D} & Fiji & N/A & 0.97 & 21.3S\\ \hline
\end{tabular}
\end{table*}

\section{Discussion} \label{sec:disc}

So can we indeed ascribe the Bolivian bolide in question CNEOS 2017-10-09 as originating from 'Oumuamua? There are 3 factors worth considering which support this proposition:

\begin{enumerate}
    \item{as discussed, the $\Delta V$ required from an object ejected from 'Oumuamua for an intercept on this date is a minimum (Figure \ref{fig:bolidedv} - red solid curve) and reaches less than 300 $\si{m.s^{-1}}$ prior to January 2016.}
    \item{'Oumuamua's light curve indicated its orientation was changing and rapidly tumbling, which many have suggested was due to a collision with another body, could this impact have occurred within our Solar System? Alternatively did 'Oumuamua fragment spontaneously due to rapidly escalating solar flux or gravitational disruption as it approached perihelion?}
    \item{it is estimated any object originating from 'Oumuamua would necessarily have to strike the Earth some time between 03:00 to 15:00 local time, this can be surmised from the animation provided at \cite{CiteAnim} (refer around October 9th). The local time of the Bolivian bolide was 08:51:48, as required.}
\end{enumerate}

A summary of the corresponding statistics for each of the 6 bolides highlighted in Figure \ref{fig:bolidezhist} is given in Table \ref{tab:loctime}. We find that the Bolivian bolide  (3rd one down) satisfies the criteria for originating from 'Oumuamua, although the Earth hyperbolic excess speed ($V_{\infty}$) is unknown due to insufficient data from CNEOS. As previously mentioned as a bare minimum this data would enable the confident rejection of a bolide as being interstellar (i.e. if $V_{\infty} < 12.3\ \si{km.s^{-1}}$), but alternatively it would potentially establish it as being interstellar using the calculations detailed in \cite{Siraj_2022}.\\

The fourth listed bolide in Table \ref{tab:loctime}, CNEOS 2017-10-15 satisfies the criterion 3 above, but is on the border-line as being excluded as interstellar due to its low $V_{\infty}$ which is 11.3 $\si{km.s^{-1}}$. It should be noted that an object originating from 'Oumuamua would likely have a high $V_{\infty} > 60\ \si{km.s^{-1}}$. We can thus confidently reject this bolide as an 'Oumuamua-related object. This leaves the last object CNEOS 2017-10-26 for which  we have no data concerning its $V_{\infty}$, which WAS observed in the time window estimated in 3 above, yet nevertheless had the second largest $\Delta V$. \\

It should be noted that there is no expected equivalence between the latitude of the observed bolide and the declination of the arrival asymptote of an incoming object, since their is uncertainty concerning two important factors:
\begin{enumerate}
    \item{ the impact parameter, b, of the object}
    \item{the $\beta$-angle the asymptote makes at its intersection with the approach $\beta$-plane centred on Earth, and the equator.}
\end{enumerate}

Both of these parameters are critical in governing this latitude of impact. \\

Assuming that CNEOS 2017-10-09 did indeed derive from 'Oumuamua we now consider the possible mass of the object. As indicated above, the velocity for this bolide is unavailable, however optical measurements determined a total radiative energy of $E_{0} = 3.0\times 10^{10}\ \si{J}$.\\

As elaborated in \cite{Brown2002-gn}, only a small fraction, $\tau$, of the total impact kinetic energy ($E$) of a meteor is converted into light. $\tau$ is dependent on many parameters whose values are generally unknown for any given object, such as the composition and porosity of the bolide in question. However $\tau$ is known to increase with the impact energy of the bolide, and therefore also its total radiative energy. A best-fit relationship between $\tau$ and $E_{0}$ from \cite{Brown2002-gn} is given as follows:
\begin{equation}\label{TAU}
\tau = a.E_{0}^{b},
\end{equation}
\begin{equation}
a = 0.1212 \pm 0.0043,\
b = 0.115 \pm 0.075,
\end{equation}

where by definition:
\begin{equation}\label{TAU2}
\tau = \frac{E_{0}}{E},
\end{equation}

Combining \ref{TAU} with \ref{TAU2} gives:
\begin{equation}\label{E}
E = c.E_{0}^{d},
\end{equation}
\begin{equation}
c = 8.2508 \pm 0.293,\
d = 0.885 \pm 0.075,
\end{equation}

Note in all the above $E$ and $E_{0}$ have units of kilotons TNT (kt).\\

For CNEOS 2017-10-09 we have $E_0 = 3.0\times 10^{10} \si{J}$ which is equivalent to $0.00717\  \si{kt}$. This gives from Equation \ref{TAU}, $\tau$ in the region of $4.6\% - 9.6\%$. This interval corresponds to a range of $E$ equivalent to $0.070 \ -\ 0.157\ \si{kt}$. We also know the expected arrival hyperbolic excess speed of the meteor w.r.t. Earth ($\sim{60}\ \si{km.s^{-1}}$), from which we can infer the possible mass of the bolide - refer Figure \ref{fig:bolivian}, where a mass $\sim{200} \si{kg}$ is indicated. This can be compared with the CNEOS 2014-01-08 bolide for which \cite{Siraj_2022} derive a mass of $\sim{460} \si{kg}$.\\

Note the above calculations of the bolide's mass suppose that 'Oumuamua is similar in composition to asteroids belonging to the Solar System and although being instructive, some scientists would doubt this supposition. This is because 'Oumuamua possessed various unusual characteristics, for instance the presence of a non-gravitational acceleration \citep{Micheli2018}, and also an unusual aspect ratio of at least 5:1, thus leading certain scientists to derive extraordinary theories as to the nature and composition of 'Oumuamua, suggesting exotic materials such as  hydrogen ice \citep{Seligman_2020}, the more mundane nitrogen ice \citep{Jackson2021,Desch2021} and even alien technology \citep{Bialy_2018}.\\

Further work might better establish the most likely cause of CNEOS 2017-10-09 being ejected from 'Oumuamua, whether via fragmentation or a collision, and also determine the most likely time of such an event.

\section{Conclusion} \label{sec:conc}

In this paper, a bolide (CNEOS 2017-10-09) possibly ejected from the object 'Oumuamua was identified from the CNEOS database of bolides, suggesting an interstellar origin for this object. Whether this attribution is valid is clearly not known with any certainty, but if it did originate from 'Oumuamua, through for instance 'Oumuamua colliding with another body prior to the ISO's discovery on 19th October 2017, or alternatively fragmenting spontaneously due to rapidly increasing solar flux or gravitational disruption, this is sympathetic with observations of 'Oumuamua's light curve which indicate a chaotic tumbling rotational state. Furthermore the bolide in question was observed on the sunlit side of the Earth in a time window in accord with an 'Oumuamua origin.\\

\bibliography{InterstellarMeteor}{}
\bibliographystyle{aasjournal}



\end{document}